# First-principles thermal equation of state and thermoelasticity of hcp Fe at high pressures

Xianwei Sha and R. E. Cohen

Carnegie Institution of Washington, 5251 Broad Branch Road, NW, Washington, D. C. 20015, U. S. A.

We investigate the equation of state and elastic properties of hcp iron at high pressures and high temperatures using first principles linear response linear-muffin-tin-orbital method in the generalized-gradient approximation. We calculate the Helmholtz free energy as a function of volume, temperature, and volume-conserving strains, including the electronic excitation contributions from band structures and lattice vibrational contributions from quasi-harmonic lattice dynamics. We perform detailed investigations on the behavior of elastic moduli and equation of state properties as functions of temperature and pressure, including the pressure-volume equation of state, bulk modulus, the thermal expansion coefficient, the Grüneisen ratio, and the shock Hugoniot. Detailed comparison has been made with available experimental measurements and theoretical predictions.

I. Introduction

Iron is one of the most abundant elements in the Earth, and is fundamental to our world. The study of iron at high pressures and high temperatures is of great geophysical interest, since both the Earth's liquid outer core and solid inner core are composed mostly of this element. Although the crystal structure of iron at the extremely high temperature (4000 to 8000 K) and high pressure (330 to 360 GPa) conditions found in the inner core is still under intensive debate,[1-10] the hexagonal-close-packed phase (ε-Fe) is commonly believed to have a wide stability field extending from deep mantle to core conditions, and serves as a starting point for understanding the nature of the inner core.[11] Significant experimental and theoretical efforts have been recently devoted to investigate various properties of hcp iron at high pressures and high temperatures. New high-pressure diamond-anvil-cell techniques have been developed or significantly improved, which makes it possible to reach higher pressures and provide more valuable information on material properties in these extreme states. First-principles based theoretical techniques have been improved in reliability and accuracy, and have been widely used to predicate the high pressure-temperature behavior and provide fundamental understandings to the experiment.

Despite intensive investigations, numerous fundamental problems remain unresolved, and many of the current results are mutually inconsistent.[11] The melting line at very high pressures has been one of the most difficult and controversial problems.[12-19] Other major problems include possible subsolidus phase transitions[2, 4, 5, 11, 20] and the magnetic structure of the dense hexagonal iron.[21-23] First-principles calculations predicted that hexagonal close-packed iron has antiferromagnetic ground state up to 50

GPa, and becomes nonmagnetic with further increase in pressure.[24, 25] Recent antiferromagnetic calculations explained the anomalous splitting of the Raman mode and the absence of hyperfine splitting in Mössbauer measurements in hpc iron.[21]

Knowledge about the elasticity of hcp iron and its pressure and temperature dependences plays a crucial role in understanding the seismological observations of the inner core, such as the low shear velocity and the elastic anisotropy. Several sets of first-principles elastic moduli have been reported for hcp Fe at high pressures[26-31], most of which are zero-temperature calculations. Steinle-Neumann et al. examined the thermoelasticity at the inner core conditions using first-principles pseudopotential calculations and the PIC model, but their calculations gave too large *c/a* ratios at high temperatures.[28] Vocadlo examined the elastic constants of iron and several iron alloys at high temperatures via *ab initio* molecular dynamics simulations and thermodynamic integration, with two selected temperatures and atomic densities for hcp Fe.[32] Here we present the calculated elasticity of nonmagnetic hcp Fe as a function of pressure and temperature using first-principles linear response calculations.

Since we focus here on iron properties at high pressures and high temperatures, we perform nonmagnetic computations. Although we provide results even at lower pressures for sake of comparison with other studies, only our results above 50 GPa, where iron is nonmagnetic according to all analyses, should be considered comparable to experiments.

There have been many discussions regarding the c/a lattice strains at high temperatures. Two earlier calculations used the particle-in-cell (PIC) model to obtain the lattice vibrational contributions, and predicted a rapid increase in the *c/a* axial ratio to above 1.7 at the core conditions.[28, 33] However, later theoretical work by Alfe and coauthors using *ab initio* molecular dynamics simulations[34-37] and experimental measurements up to 2000K[12] both gave much smaller temperature dependences of the *c/a* ratio. We found that the results from the first-principles linear response calculations and the PIC model usually agree well except when the lattice approaches instability, and both theoretical techniques predicted a slight increase in the axial ratio with temperature, in contradiction to the earlier PIC computations.[38] Since the on-site anharmonicity in nonmagnetic hcp Fe is small up to the melting temperature,[38] here we use first-principles linear response calculations based on the full-potential linear-muffin-tin-orbital (LTMO) method and quasi-harmonic approximation to examine the thermal equation of state of nonmagnetic hcp Fe.[39]

In section II we detail the theoretical methods to perform the first-principles calculations and obtain the thermal properties and elastic moduli. We present the results and related discussions about the thermal equation of state in section III, and about the thermoelasticity in section IV. We conclude with a brief summary in Section V.

## II.    Theoretical methods

The Helmholtz free energy F for many metals has three major contributions[40]

$$F(V,T,\delta)=E_{static}(V,\delta)+F_{el}(V,T,\delta)+F_{ph}(V,T,\delta) \qquad (1)$$

with V as the volume, T as the temperature, and $\delta$ as the strain. $E_{static}$ is the zero-temperature energy of a static lattice, $F_{el}$ is the thermal free energy arising from electronic excitations, and $F_{ph}$ is the lattice vibrational energy contribution. We obtain both $E_{static}$ and $F_{el}$ from first-principles calculations directly, assuming that the eigenvalues for given lattice and nuclear positions are temperature-independent and only the occupation numbers change with temperature through the Fermi-Dirac distribution.[33, 38, 41] The validity of the static eigenvalue approximations is well justified by the fact that the calculated electronic entropies of nonmagnetic hcp Fe agree well with the values from self-consistent high temperature Linear-Augmented-Plane-Wave (LAPW) method[33] over a wide temperature (6000-9000K) and volume (40-90 bohr$^3$/atom) range. The linear response method gives the phonon dispersion spectrum and phonon density of states, which provide both a microscopic basic for and a means of calculating the thermodynamic and elastic properties.[11] We obtain the vibrational free energy within the quasiharmonic approximation.

The computational approach is based on the density functional theory and density functional perturbation theory, using multi-κ basis sets in the full-potential LMTO method.[42, 43] The induced charge densities, the screened potentials and the envelope functions are represented by spherical harmonics inside the non-overlapping muffin-tin spheres surrounding each atom and by plane waves in the remaining interstitial region. We use the Perdew-Burke-Ernzerhof (PBE) generalized-gradient-approximation (GGA) for the exchange and correlation functional.[44] The **k**-space integration is performed over a 12×12×12 grid using the improved tetrahedron method.[45] We use the perturbative approach to calculate the self-consistent change in the potential,[46, 47] and determine the dynamical matrix for a set of irreducible **q** points on a 6×6×6 reciprocal lattice grid. Careful convergence tests have been made against **k** and **q** point grids and other parameters. We examine hcp Fe at volumes from 40 to 80 bohr$^3$/atom and at c/a ratios from 1.5 to 1.7 in 0.05 interval. We determine the equilibrium thermal properties by minimizing the Helmholtz free energies with *c/a* ratio at a given temperature and volume.

We obtain the elastic moduli as the second derivatives of the Helmholtz free energies with respect to strain tensor, by applying volume-conserving strains and relaxing the symmetry-allowed internal coordinates. For hexagonal crystals, the bulk modulus K and shear modulus $C_S$ yield the combinations of the elastic moduli

$$K = [C_{33}(C_{11} + C_{12}) - 2C_{13}^2]/C_S \qquad (2)$$

$$C_S = C_{11} + C_{12} + 2C_{33} - 4C_{13} \qquad (3)$$

To make direct comparisons to the ultrasonic measurements, we use the adiabatic bulk modulus $K_s$[40]

$$K_S = (1 + \alpha \gamma T) \times K_T , \qquad (4)$$

where $K_T$ is the isothermal bulk modulus, $\alpha$ is the thermal expansivity, and $\gamma$ is the Grüneisen parameter. We obtain the equation of state parameters $K_T$, $\alpha$, and $\gamma$ as functions of temperature and pressure from the first-principles linear response calculations.

We calculate $C_S$ by varying the *c/a* ratio at a given volume:

$$\varepsilon(\delta) = \begin{pmatrix} \delta & 0 & 0 \\ 0 & \delta & 0 \\ 0 & 0 & (1+\delta)^{-2} - 1 \end{pmatrix}, \qquad (5)$$

where $\delta$ is the strain magnitude. The Helmholtz free energy $F(\delta)$ is related to $\delta$ as:

$$F(\delta) = F(0) + C_s V \delta^2 + O(\delta^3), \qquad (6)$$

with F(0) as the free energy of the unstrained structure.

The volume dependences of the equilibrium *c/a* ratio are related to the difference in the linear compressibility along the *a* and *c* axes

$$-\frac{d\ln(c/a)}{d\ln V} = (C_{33} - C_{11} - C_{12} + C_{13})/C_S \ . \tag{7}$$

We apply a volume-conserving orthorhombic strain to calculate the difference between $C_{11}$ and $C_{12}$, $C_{11}-C_{12}=2C_{66}$,

$$\varepsilon(\delta) = \begin{pmatrix} \delta & 0 & 0 \\ 0 & -\delta & 0 \\ 0 & 0 & \delta^2/(1-\delta^2) \end{pmatrix} \ . \tag{8}$$

The corresponding free energy change is:

$$F(\delta) = F(0) + 2C_{66}V\delta^2 + O(\delta^4) \ . \tag{9}$$

We use a monoclinic strain to determine $C_{44}$

$$\varepsilon(\delta) = \begin{pmatrix} 0 & 0 & \delta \\ 0 & \delta^2/(1-\delta^2) & 0 \\ \delta & 0 & 0 \end{pmatrix}, \tag{10}$$

which leads to the energy change

$$F(\delta) = F(0) + 2C_{44}V\delta^2 + O(\delta^4) \ . \tag{11}$$

When evaluating $C_{44}$ and $C_{66}$, we relax the internal degree of freedom by minimizing the total energy with respect to the atomic positions in the two atom primitive unit cell.[24, 29] Since the leading error term is third order in δ for $C_s$ and fourth order for $C_{44}$ and $C_{66}$, we include both positive and negative strains to calculate $C_s$. We choose 4-6 values for each strain ranging from 0 to 0.03, and perform first-principles linear response calculations to obtain the band structure and phonon density of states for all the strained structures at each volume. We then calculate the Helmholtz free energies at temperatures from 0 to 6000 K, and fit a polynomial of the free energies to the strain magnitudes. The quadratic coefficients of the polynomial fitting give the elastic moduli that appear in the equations of motion and directly give sound velocities.[24, 48, 49]

## III. Thermal equation of state

We present in Fig. 1 the calculated phonon density of states (DOS) of hcp Fe at the c/a ratio of 1.6 and volumes of 40, 60, and 70 bohr$^3$/atom. Nuclear resonant inelastic x-ray scattering techniques have been used to measure the phonon DOS of hcp Fe up to high pressures,[50-53] and our first-principles linear response results agree well with the experimental measurements. The Raman-active $E_{2g}$ phonon correlates with the zone–edge acoustic mode, the elastic modulus $C_{44}$, and shear-wave velocity, and their frequencies at high pressures have been recently measured using Raman spectroscopy.[54, 55] Our linear-response $E_{2g}$ frequencies show excellent agreement with the first-principles frozen-phonon values[21, 54] at both ambient and high pressures, as shown in Fig. 2. Although theory gives similar pressure dependences of the Raman frequencies as experiment,[54, 55] all the theoretical calculations overestimate the $E_{2g}$ frequencies by ~15 %. At low pressures, the antiferromagnetic nature of the ground state hcp Fe leads to splitting of the Raman frequencies,[21] and substantial temperature and compositional dependence.[55] All these account for some of the discrepancies between theory and experiment.

We fit the calculated Helmholtz free energies at each given temperature to an equation of state (EoS) formulation to obtain the bulk modulus and thermal pressures. Due to its versatility and high accuracy, we choose the Vinet EoS form[56-58]

$$F(V,T) = F_0(T) + \frac{9K_0(T)V_0(T)}{\xi^2}\{1 + \{\xi(1-x) - 1\}\exp\{\xi(1-x)\}\}] \quad (12)$$

where x = $(V/V_0)^{1/3}$, $K_0(T)$ is the bulk modulus, $\xi = \frac{3}{2}(K_0' - 1)$ and $K_0' = [\frac{\partial K(T)}{\partial P}]_0$.

The subscript 0 throughout represents the standard state P= 0 GPa. We list the calculated Vinet EoS parameters at ambient condition in Table I. The current LMTO results agree well with recent first principles calculations for nonmagnetic hcp Fe using the all-electron Linearized-Augmented-Plane-Wave (LAPW) Method[24] and the Projector-Augmented-Wave (PAW) method,[59] both using the PBE GGA functional. The discrepancy between the nonmagnetic calculations and diamond-anvil-cell experiments[60-62] is significantly larger for hcp Fe than for typical transition metals. As shown in earlier calculations, including the antiferromagnetic ground state in the first-principles calculations helps to significantly improve the agreements with experiment at low pressures (P < 50 GPa).[24] The temperature dependences of the Vinet EoS parameters $V_0(T)$, $K_0(T)$, and $K_0'(T)$ are plotted in Fig. 3, which show typical features of transition metals: thermal expansion and decrease of bulk modulus with increasing temperature.[41, 63]

We obtain the pressure analytically from the Vinet EoS parameters:

$$P(V,T) = \{\frac{3K_0(T)(1-x)}{x^2}\} \exp\{\xi(1-x)\} \qquad (13)$$

In Fig. 4 we show the calculated pressure-volume equation of state for hcp Fe at temperatures between 0 to 3000 K in 500 K intervals. Compared to the ambient-temperature x-ray diffraction measurements, our first-principles results agree well with the experiments to 78 GPa with Ar and Ne pressure-transmitting media[64] and to 304 GPa without a medium.[60] The discrepancies between the calculated and experimental data are larger at low pressures (< 50 GPa) mainly due to the neglect of magnetism in the calculations, and spin-polarized GGA calculations of an antiferromagnetic structure agree better with the experiment.[24]

We obtain the thermal pressures as functions of volume and temperature according to the pressure differences among the EoS isotherms, as shown in Fig. 5. The thermal pressures are small and essentially volume-independent at low temperatures, but increase dramatically and show complex volume-dependence at high temperatures. At a given volume, the thermal pressures increase linearly with temperature. All these are similar to the behavior previously reported in bcc Fe[41] and Ta.[65]

In order to more accurately extract higher order derivatives, we fit a Debye model with a Debye temperature $\theta_D(V,T)$, which is a function of volume and temperature. Such a model for the free energy does not assume that the phonon spectrum is Debye-like, and has been successfully used for many complex minerals.[66-70] An accurate high-temperature global equation of state can be formed from the 0 K Vinet isotherm plus a volume-dependent thermal free energy $F_{th}$,[40]

$$F_{th} = RT[\frac{9}{8}(\frac{\theta_D}{T}) + 3\ln(1 - e^{-\theta_D/T}) - D(\frac{\theta_D}{T})] \quad , \tag{14}$$

where the Debye function $D(\theta_D/T)$ is defined as

$$D(\frac{\theta_D}{T}) = 3(\frac{T}{\theta_D})^3 \int_0^{\theta_D/T} \frac{z^3 dz}{e^z - 1} \quad . \tag{15}$$

We find the Debye temperature function $\theta_D(V,T)$ at 0K by numerical integration of the low-frequency part of the phonon density of state, and solve Eqn. 14 to obtain $\theta_D(V,T)$ at other temperatures. The calculated and fitted thermal free energies agree well at different temperatures and volumes with an rms deviations of ~0.4 mRy.

We derive various thermal equation of state properties analytically from the Helmholtz free energy.[40] The thermal expansion coefficient α is:

$$\alpha = -\frac{1}{V}(\frac{\partial^2 F}{\partial T \partial V})/(\frac{\partial^2 F}{\partial V^2})_T = \frac{3R\gamma_D}{K_T V}[4D(\frac{\theta}{T}) - \frac{3(\theta/T)}{e^{\theta/T} - 1}] \tag{16}$$

The calculated α increases linearly with temperature at both ambient and high pressures (Fig. 6). The calculations show fair agreements with the shock wave[71] and in situ x-ray[61] measurements at high pressures and temperatures (P ≈ 200 GPa, T ≈ 5200 K). Isaak and Anderson estimated the thermal expansivity of hcp Fe at high pressures and temperatures based on thermodynamic analysis of compression curves constructed from ultrasonic elasticity, static compression, and shock compression and temperature measurements.[72] Compared to their high pressure and high temperature data, our first-principles calculations give better agreements with the shock and in situ x-ray measured data.

The Anderson-Grüneisen parameter $\delta_T$ is used to characterize the pressure dependence of the thermal expansion coefficient:

$$\delta_T = (\frac{\partial \ln \alpha}{\partial \ln V})_T = -\frac{1}{\alpha K_T}(\frac{\partial K_T}{\partial T})_P \tag{16}$$

The calculated $\delta_T$ drops rapidly with increasing pressure at a given temperature, and shows complex temperature dependences, as shown in Fig. 7(a). For many materials, the parameter $\delta_T$ has been parameterized as a function of volume:[40]

$$\delta_T = \delta_T(\eta = 1) \times \eta^\kappa \tag{17}$$

where $\eta = V/V_0(T_0)$. The equation works well for transition metals such as bcc Ta[65] and metal oxides such as MgO.[73] As shown in Fig. 7(b), although $\delta_T$ of hcp Fe shows a strong decrease during compression, it does not drop as rapidly as power order at high pressures, similar to what has been observed in bcc Fe.[41]

The Grüneisen ratio γ is an important parameter in understanding the relationship between the thermal and elastic properties:

$$\gamma = V(\frac{\partial P}{\partial U})_V = \frac{\alpha K_T V}{C_V} = V \frac{\partial^2 F}{\partial V \partial T} \bigg/ (\frac{\partial U}{\partial T})_V \tag{18}$$

where U is the internal energy. Many different techniques have been used to determine the Grüneisen ratio of hcp Fe, including nuclear resonant inelastic x-ray scattering,[51] Raman,[54] x-ray diffraction,[61, 74] shock wave,[75] and thermodynamic analysis.[76, 77] At a given pressure, our calculated γ first increases with temperature, and then drops rapidly at high temperatures (T >1500 K), as shown in Fig. 8. The pressure dependence of γ is complex and strongly temperature dependent. The calculated ratios at 500K agree fairly with ambient-temperature x-ray diffraction measurements. The volume dependence of the Grüneisen ratio is defined by the parameter $q$:

$$q = \frac{\partial \ln \gamma}{\partial \ln V} \quad . \tag{19}$$

The parameter $q$ is usually treated as a constant, and its experimental value for hcp Fe varies from 0.6 to over 1.6 depending on the pressure range and measuring methods.[11] Our calculations show that $q$ strongly depends on both the temperature and pressure, and even becomes negative at some pressure and temperature regimes [Fig. 9]. Similar complex behavior of parameter $q$ was previously reported for bcc Ta[65] and bcc Fe[41].

Shock compression data gives the high-pressure high-temperature equation of state along the shock Hugoniot. We calculate the relationship between the pressures $P_H$ and temperatures $T_H$ along the Hugoniot according to the Rankine-Hugoniot equation:[40]

$$\frac{1}{2} P_H [V_0(T_0) - V] = E_H - E_0(T=0) \tag{20}$$

E is the internal energy. We obtain $P_H$ and $T_H$ based on our thermal equation of state results by varying the temperature at a given volume until the Rankine-Hugoniot equation is satisfied. The calculated data agree well with the experimental data for both the shock Hugoniot[78] and the temperatures along the Hugoniot,[79] and also show good agreements with earlier thermodynamic estimations using plausible bounds for specific heat and experimental constraints for the Grüneisen parameter,[80] as shown in Fig. 10.

IV.    Thermoelasticity

Many different sets of experimental[81-89] and theoretical[24, 26, 28-30, 59, 90-94] elastic moduli at ambient or zero temperatures have been reported for ε-Fe. We present our calculated static moduli of nonmagnetic hcp Fe as a function of volume in Fig. 11, in comparison to some available experimental and theoretical data. One of the major reasons for the wide distribution of the experimental data is because single crystal samples are not available for hcp Fe. The single-crystal elastic moduli extracted from radial x-ray diffraction data on polycrystalline samples under nonhydrostatic compression contain large errors, since the assumption of a single uniform macroscopic stress applied to all grains is violated due to plastic deformation.[95, 96] As shown for hcp cobalt, the $C_{11}$, $C_{33}$, $C_{12}$ and $C_{13}$ obtained from polycrystalline samples are 20% off with respect to single-crystal measurements, and the discrepancies are up to 50% and 300% for shear moduli $C_{66}$ and $C_{44}$.[95] Our calculated elastic moduli show a strong increase with pressure, and agree fairly with experiment and earlier theoretical calculations. The large discrepancies between theory and experiment at low pressures are attributed to the anti-ferromagnetic ground state of hcp Fe, which is predicted to vanish at pressures higher than 50 GPa.[21]

Most of the experiments only give the elastic moduli at ambient temperature, and only recently has it become possible to examine the temperature effects on sound velocities using nuclear inelastic x-ray scattering in a laser-heated diamond anvil cell.[85] In Fig. 12 we show our calculated elastic moduli as a function of temperature at several different volumes, in comparison to previous theoretical results obtained using a plane wave mixed basis method and PIC model.[28] At a given atomic volume, our calculated elastic moduli show modest linear changes with the temperature. Most of the moduli show different temperature dependences than obtained by Steinle-Neumann et al.,[28] due to the large c/a ratios at high temperatures obtained in that study. Vocadlo reported that the elastic moduli of Fe and iron alloys do not show any significant variation with temperature at a given atomic density,[32] similar to what we observe for hcp Fe here. We interpolate our high-pressure high-temperature moduli to obtain the elastic properties at the two temperatures that Vocadlo examined for ε-Fe. $C_{13}$ and $C_{33}$ agree well in ~5%, and $C_{11}$ and $C_{12}$ agree within ~10%. However, the differences between the predicted shear moduli $C_{66}$ and $C_{44}$ are large, 15% and 35%, respectively. Our zero-temperature shear moduli agree well with Vocadlo's earlier work[30], so the differences come from the thermal contributions. We use linear response lattice dynamics and quasi-harmonic approximations, and Vocadlo used *ab initio* molecular dynamics and thermodynamic integration to obtain the thermal contributations. As shown in earlier calculations using both thermodynamic integration and the PIC model,[34, 38] the on-site anharmonicity in ε-Fe is small up to the melting temperature. The discrepancies might also come from the different set-ups in the first-principles calculations. Vocadlo used a 64-atom supercell and 4 irreducible *k* points in her *ab initio* molecular dynamics simulations. We carefully compare the calculated $C_{44}$ values at different *k* point meshes up to 24×24×24 and *q* meshes up to 6×6×6, and make sure our results are converged. Further experimental information is needed to validate these first principles data.

The calculated high-pressure high-temperature elastic moduli of hcp Fe can be used to calculate the sound velocity of the compressional and shear waves for hcp-Fe at extreme states, including under the Earth's core conditions, which could be directly compared with seismic wave measurements and help to understand the origin of elastic anisotropy of the Earth's inner core.

## V.     Conclusions

In summary, we present the thermal equation of state properties and thermoelasticity of nonmagnetic hcp Fe at high pressures from first-principles linear response calculations. The calculated lattice dynamics at high pressures agrees with nuclear resonant inelastic x-ray scattering and Raman measurements. The calculated pressure-volume equation of state, the thermal expansion coefficient at high pressures and temperatures, Grüneisen ratio, and shock Hugoniot all show fair agreements with available experimental data. Deviations from experiment are probably due to errors in DFT, rather than our methodology, and show the need for inclusion of magnetic fluctuations, such as through DFMT.[97,98] The variation of the Gruneisen parameter with volume, given by the parameter $q$, which is usually considered as a constant, shows strong temperature and pressure dependences in our calculations. The calculated static elastic and bulk moduli at ambient temperature are in fairly good agreements with measurements and previous calculations. At a given atomic volume, the elastic moduli show modest linear changes with temperature. This is the most comprehensive study of elasticity and equation of state of high pressure iron yet done, and should provide constraints on anisotropy and thermal behavior of iron under extreme conditions, such as in Earth's core, and provides constraints on understanding the seismology of the Earth's inner core.

**Acknowledgements**

We thank S. Y. Savrasov for kind agreement to use his LMTO codes and many helpful discussions. This work was supported by DOE ASCI/ASAP subcontract B341492 to Caltech DOE w-7405-ENG-48 and by NSF EAR-0738061, and the Carnegie Institution of Washington. Computations were performed at the Geophysical Laboratory and on ALC at Lawrence Livermore National Lab.

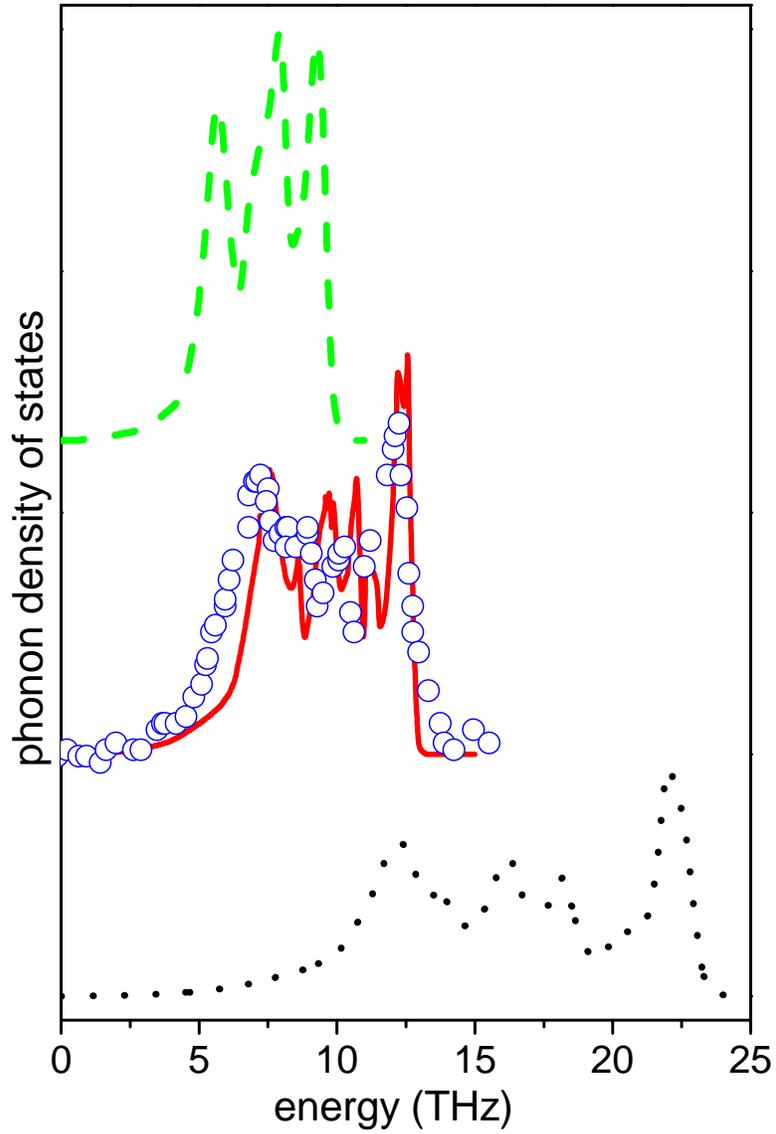

Fig. 1 The calculated phonon density of states for nonmagnetic hcp Fe at *c/a* ratio of 1.6 and volumes of 40, 60 and 70 bohr$^3$/atom, shown as the dotted, solid and dashed lines, respectively. The computed data at 60 bohr$^3$/atom agree with the nuclear resonant inelastic x-ray scattering measurements at 50 GPa (dots, ref. 50), where the sample has a similar density according to the experimental pressure-volume equation of state.

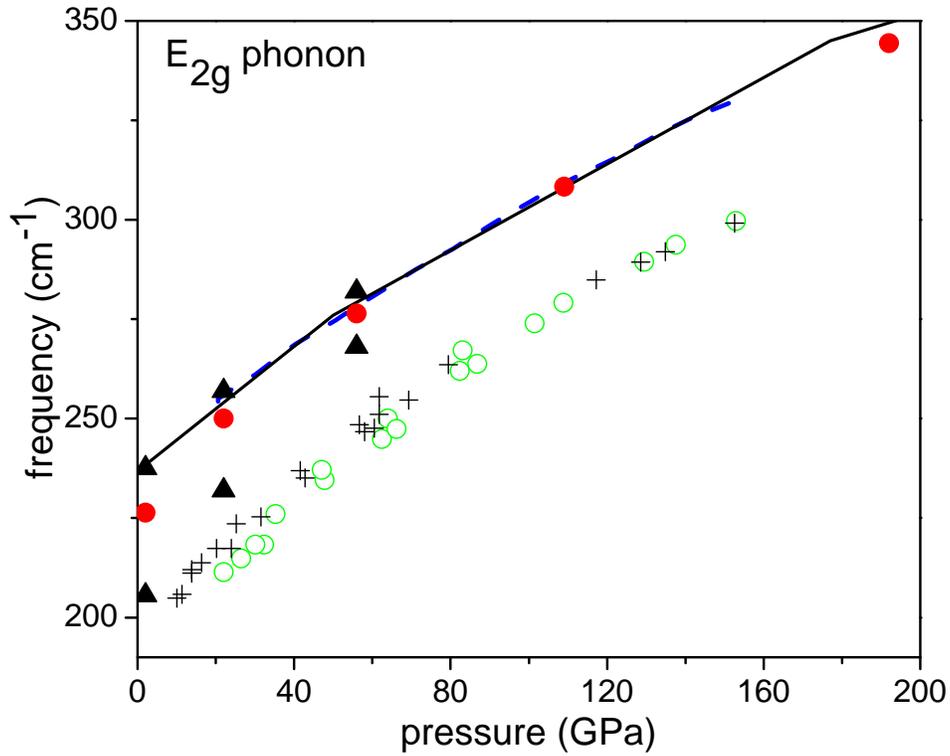

Fig. 2 The pressure dependence of the $E_{2g}$ phonon frequencies for hcp Fe. The linear-response data (solid line) agree well with frozen-phonon calculations (dotted line, ref. 54; filled circles, ref. 21), both assuming nonmagnetic hcp phase. Results from antiferromagnetic theoretical calculations (filled triangles, ref. 21) and Raman measurements (open circles, ref. 54; cross, ref. 55) are also shown.

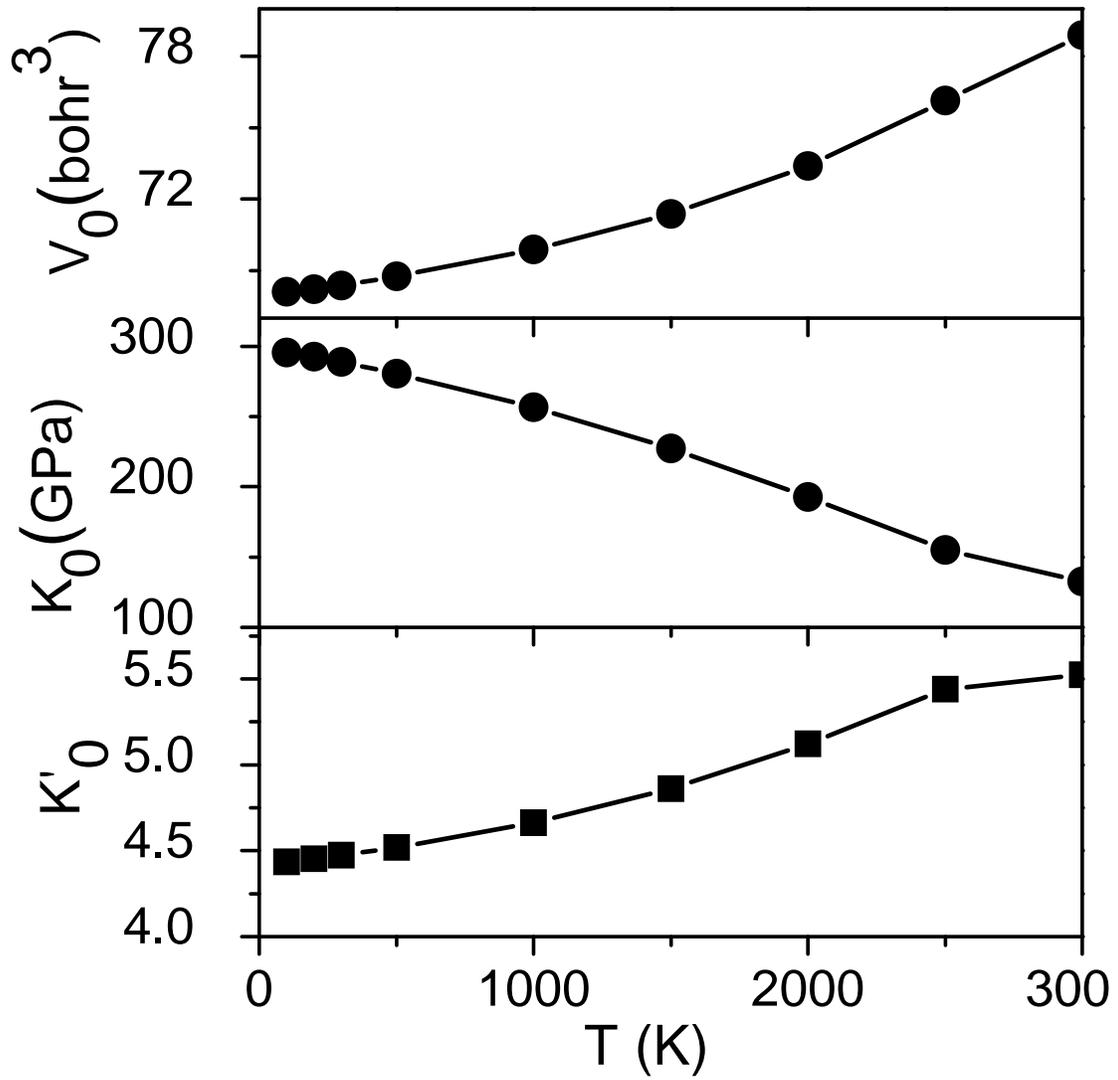

Fig. 3 The fitted Vinet equation of state parameters $V_0(T)$, $K_0(T)$, and $K_0'(T)$ as functions of temperature for nonmagnetic hcp Fe.

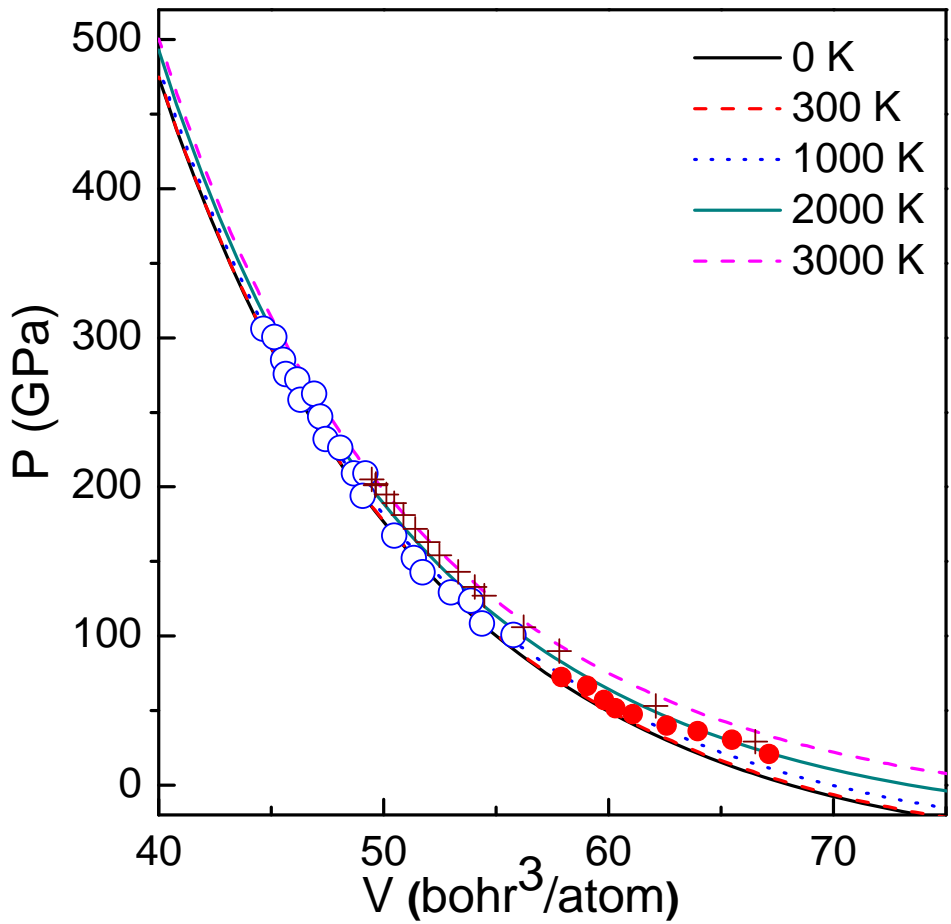

Fig. 4 The calculated pressure–volume equation of state (lines) for hcp Fe at several selected temperatures. The ambient-temperature results agree with the diamond-anvil-cell x-ray diffraction measurements (filled circles, ref. 64; open circles, ref. 60; cross, ref. 61).

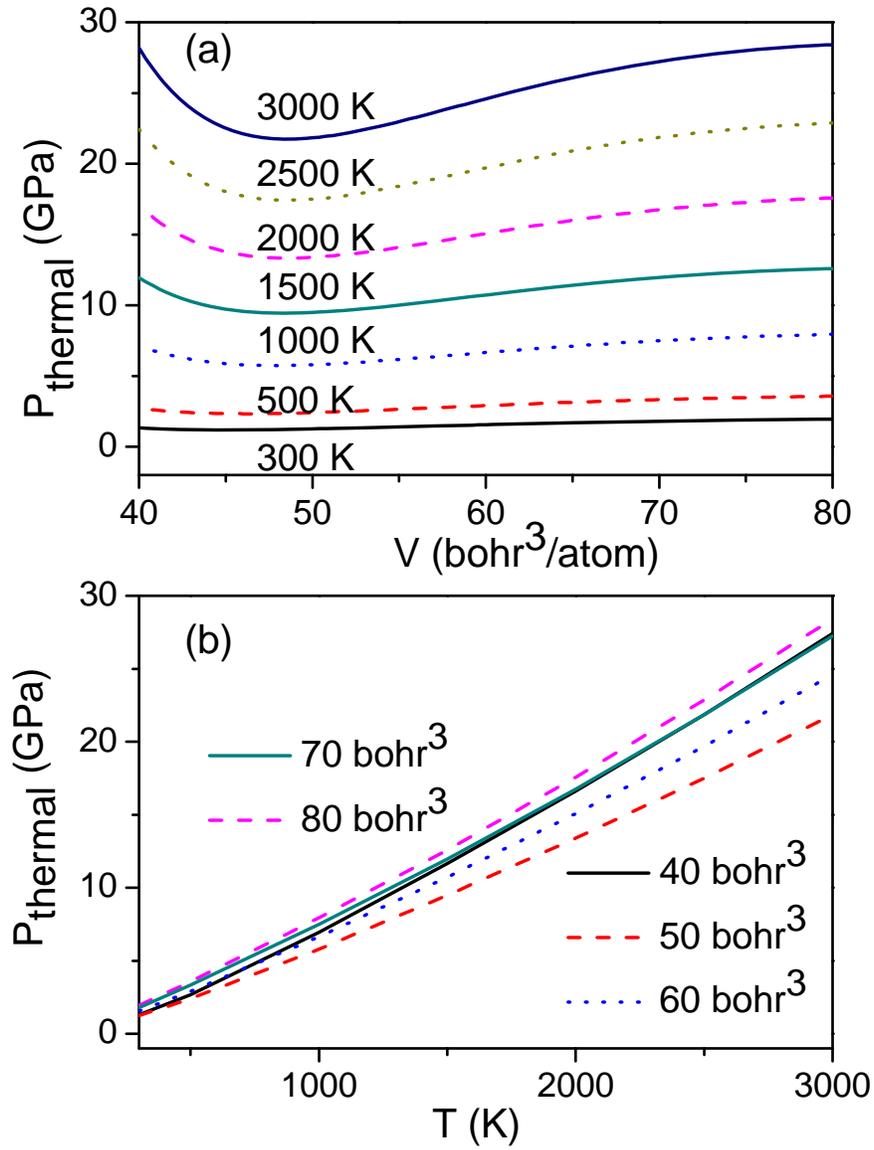

Fig. 5 The calculated thermal pressures of hcp Fe as functions of volume at several selected temperatures (a), and as functions of temperature at selected volumes (b).

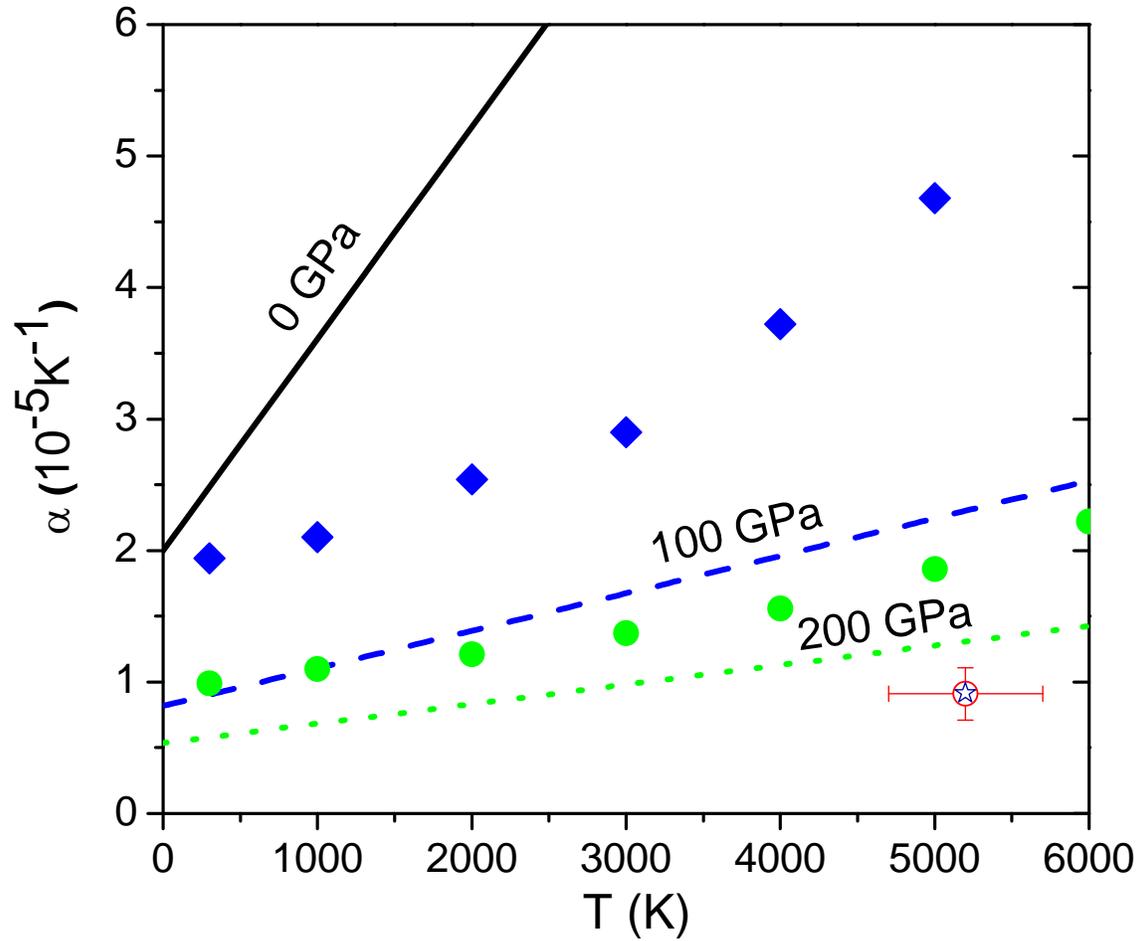

Fig. 6 The calculated thermal expansion coefficients of hcp Fe as functions of temperature at 0, 100 and 200 GPa, shown as the solid, dashed and dotted lines, in comparison to the shock compression data at 202±3 GPa (open circle with error bar, ref. 66), in situ x-ray measurement at 202 GPa (star, ref. 61), and estimated values at 100 GPa (filled diamonds, ref. 67) and 200 GPa (filled circles, ref. 67) based on thermodynamic analysis of compression curves constructed from ultrasonic elasticity, static compression, and shock compression and temperature measurements.

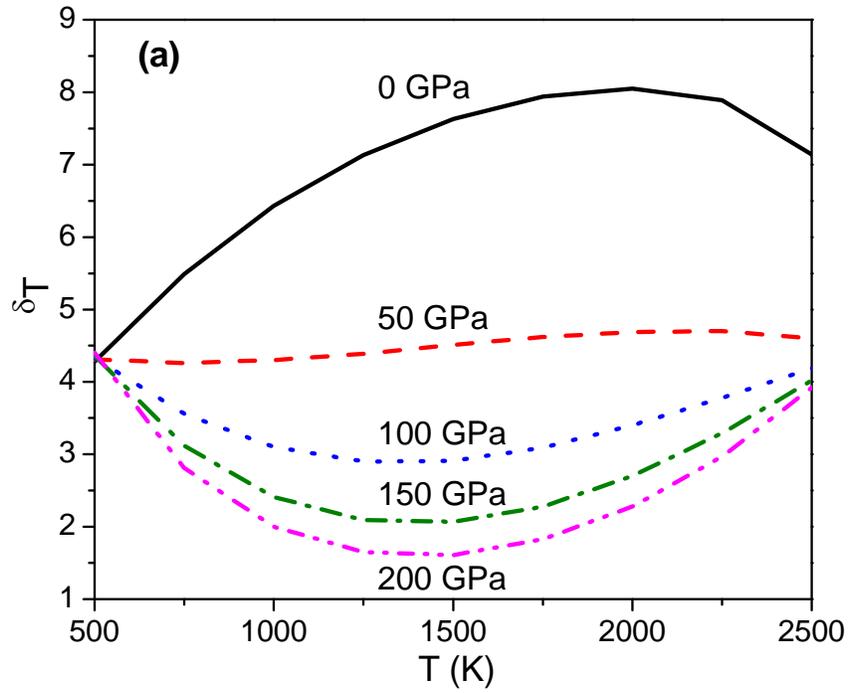

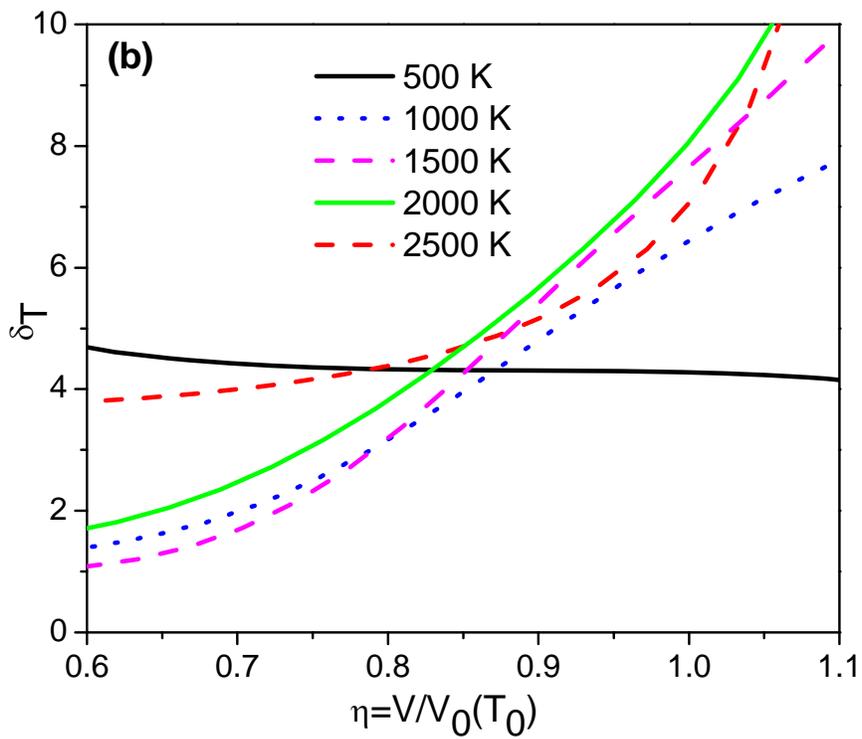

Fig. 7 The Anderson-Grüneisen parameter $\delta_T$ as functions of temperature (a) and volume (b).

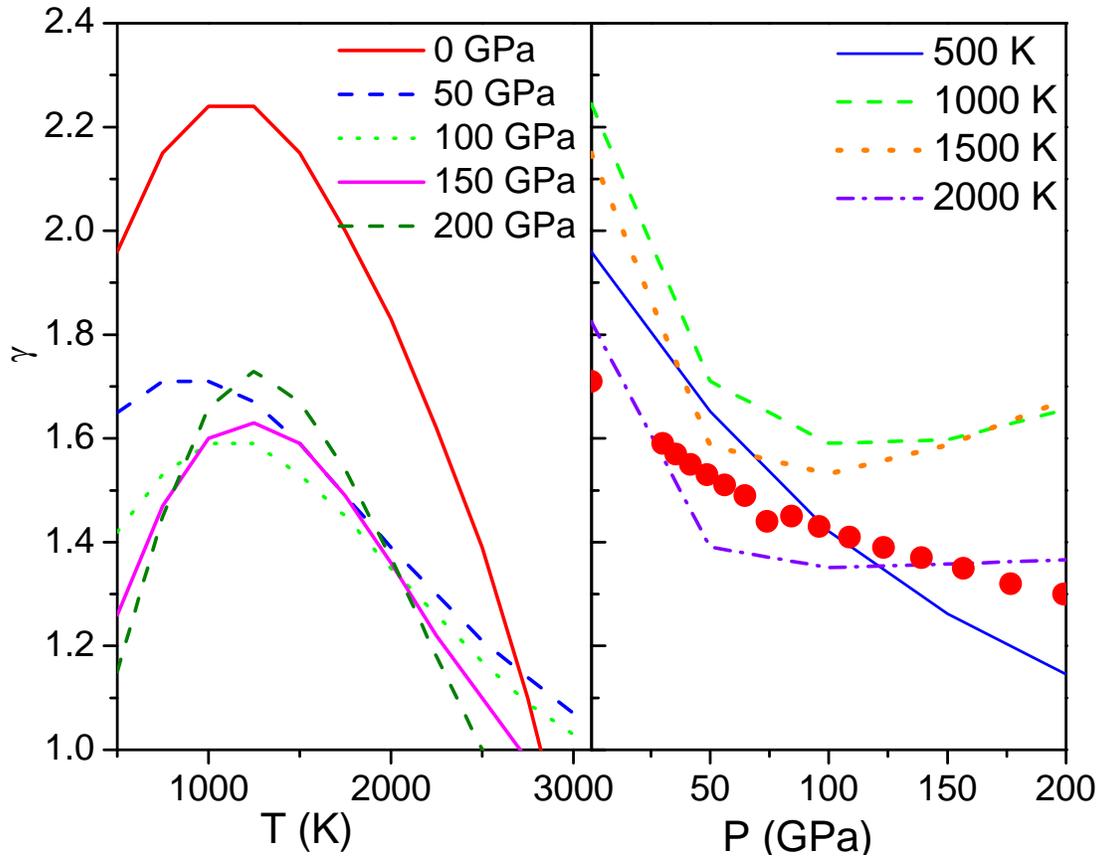

Fig. 8 The Grüneisen ratio γ of hcp Fe as functions of temperature (a) and pressure (b). The ambient-temperature x-ray diffraction data (filled circles, refs. 61 and 69) are also shown.

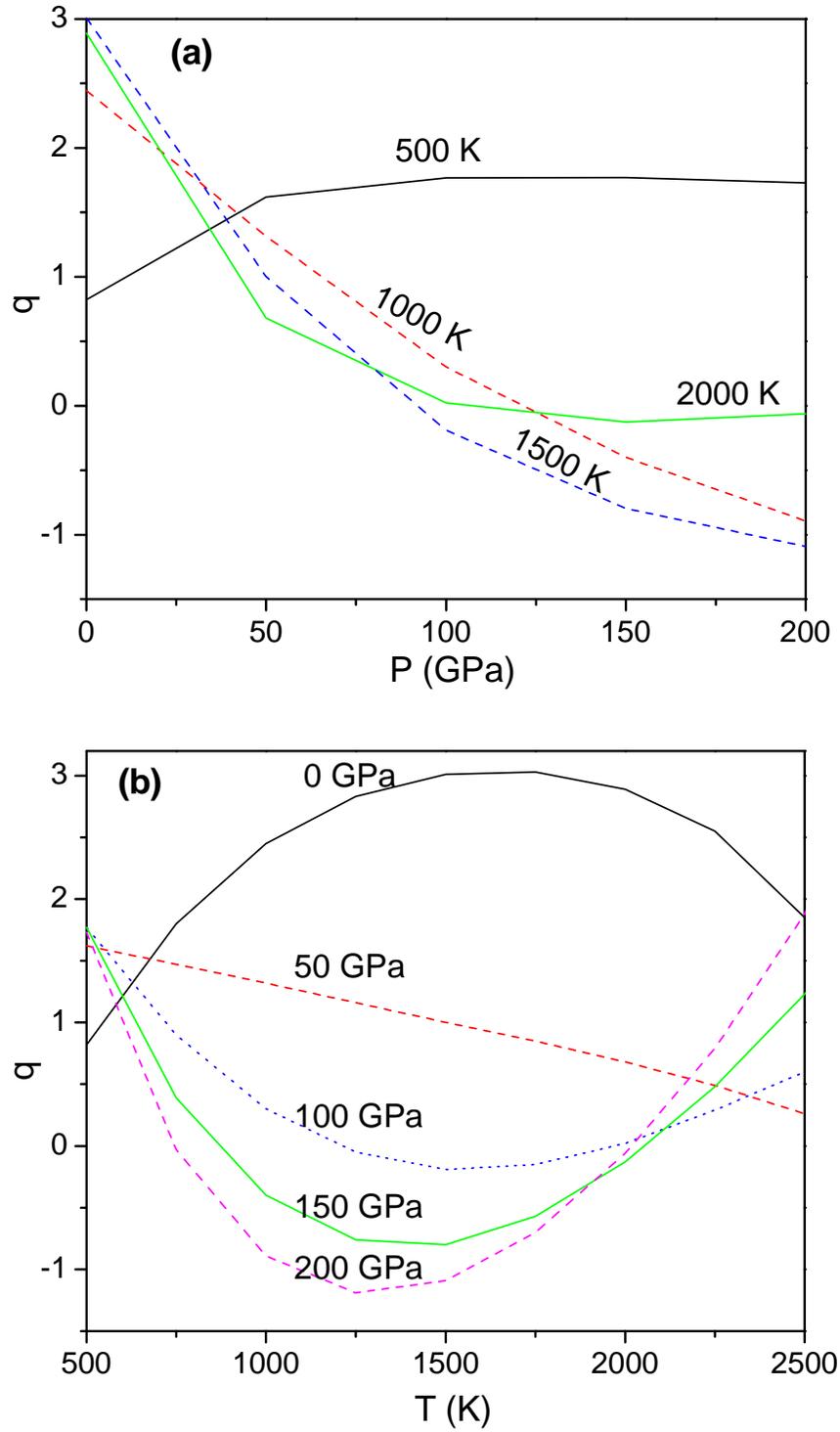

Fig. 9 The pressure (a) and temperature (b) dependences of the parameter *q* for hcp Fe.

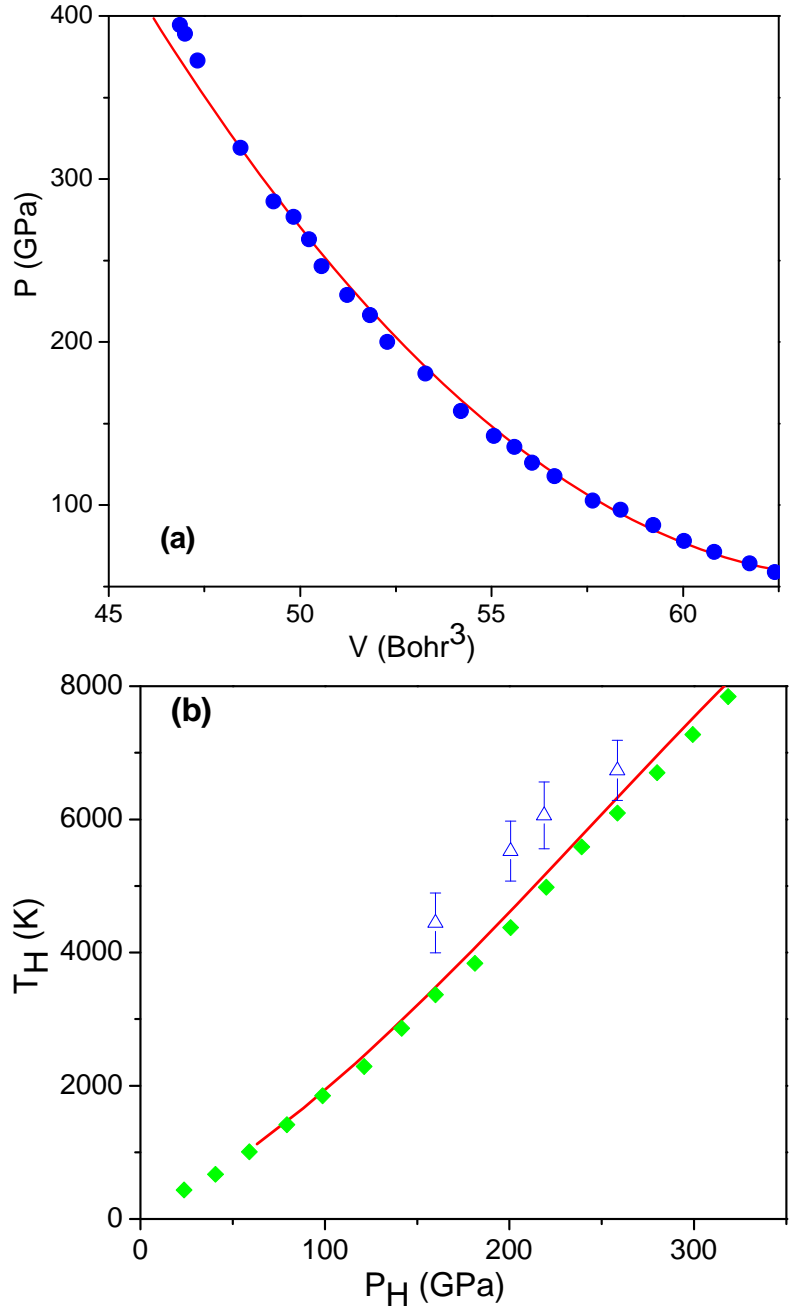

Fig. 10 Shock Hugoniot (a) and the temperatures along the Hugoniot (b) for hcp Fe. First-principles calculated data are denoted as lines, in comparison to the shock experimental data (filled circles, ref. 73; open triangles with error bars, ref. 74) and previous theoretical Hugoniot temperatures (filled diamonds, ref. 75).

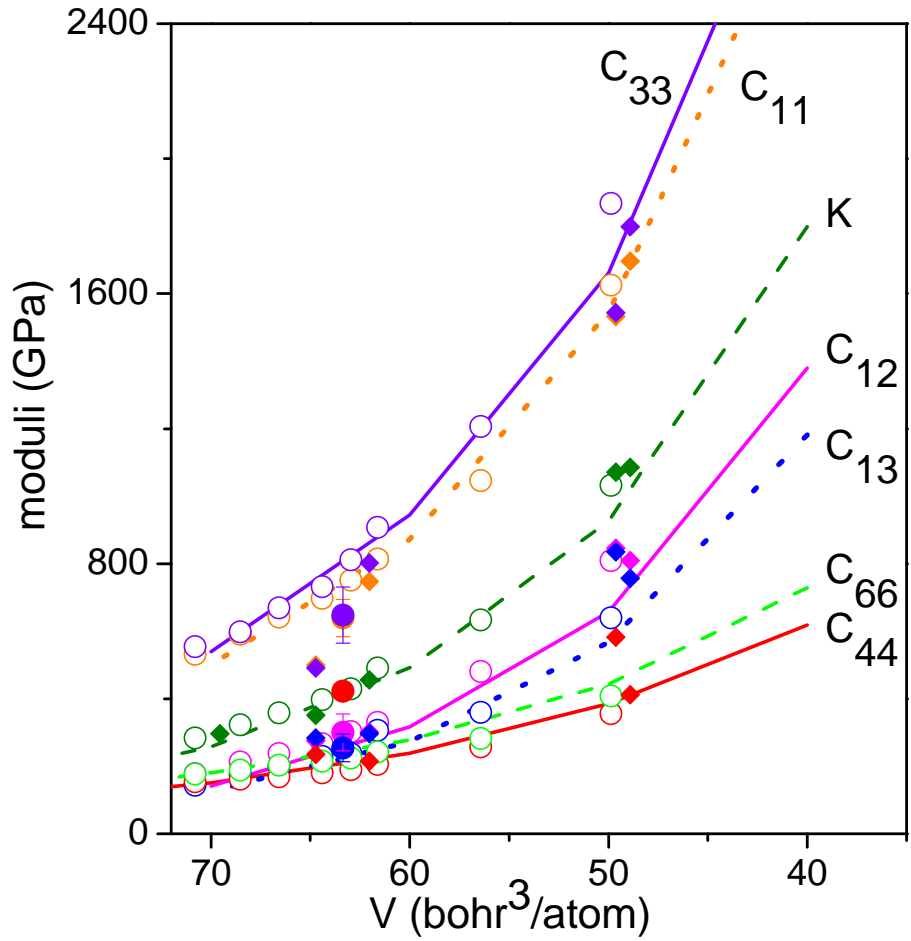

Fig. 11 Static elastic and bulk moduli of hcp Fe (lines) as functions of atomic volume, in comparison to the augmented-plane-wave plus local orbital calculated results (open circles, ref. 24), and ambient-temperature X-ray diffraction and ultrasonic experimental data (filled diamonds, refs. 81 & 82; filled circles with error bars, ref. 83).

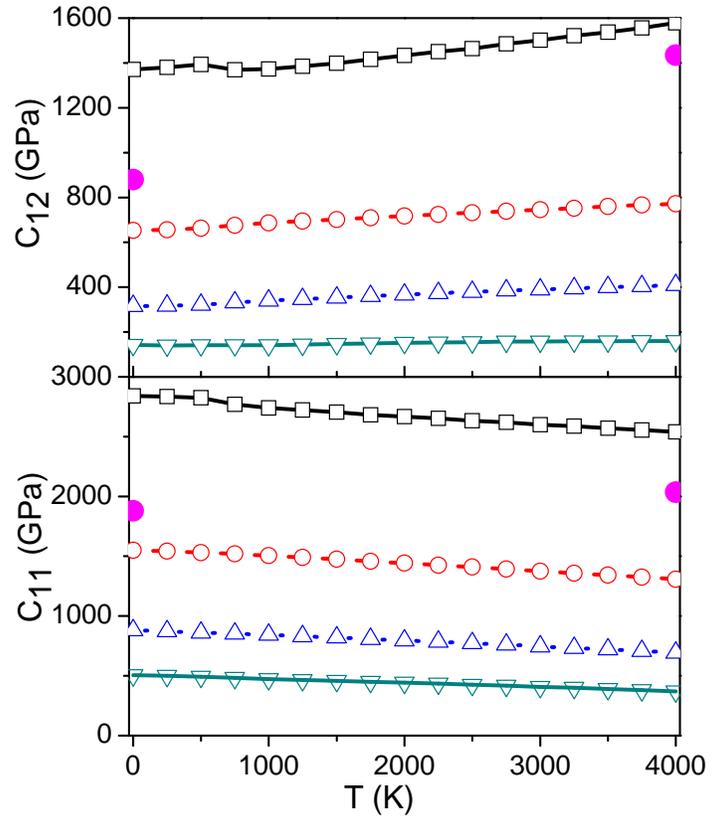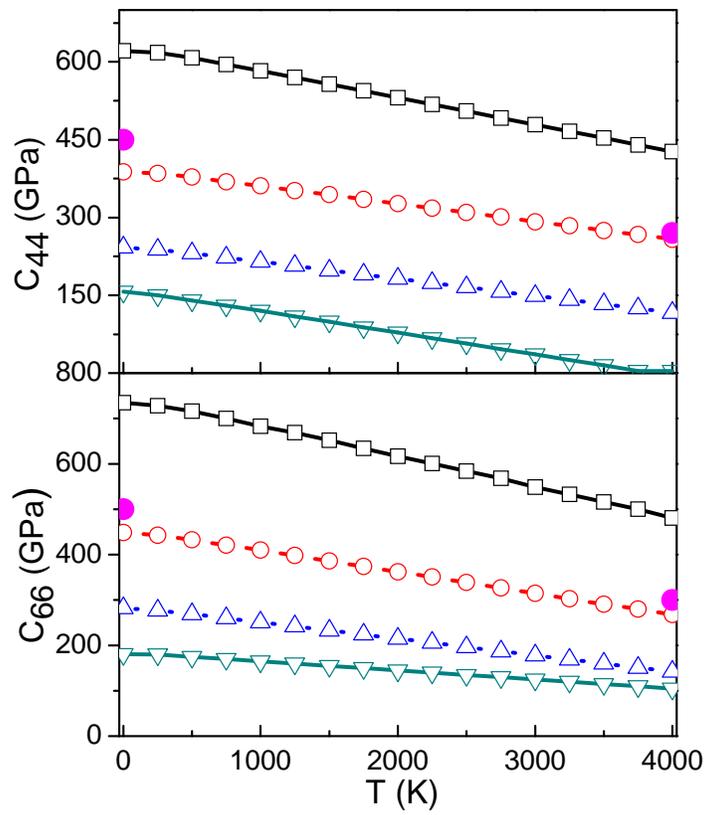

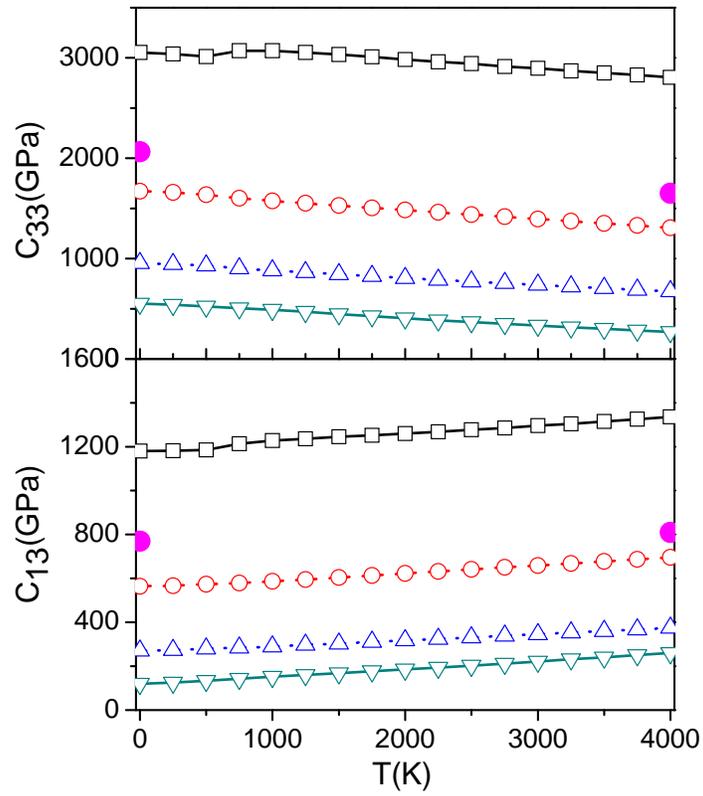

Fig. 12 The calculated temperature dependences of the elastic moduli for nonmagnetic hcp Fe, at volumes from 40 (uppermost curve) to 70 bohr$^3$/atom (lowest curve) in 10 bohr$^3$/atom interval. Previous first-principles results using a plane-wave mixed basis method and PIC model at 48 bohr$^3$/atom (filled circles, ref. 28) are also shown.

Table I  The equation-of-state parameters for hcp Fe. All the theoretical calculations are performed on nonmagnetic hcp Fe, except two antiferromagnetic configurations denoted as afmI and afmII.

|  | $V_0$ (bhor$^3$) | $K_0$ (GPa) | $K_0$' |
|---|---|---|---|
| this study | 68.1 | 296 | 4.4 |
| expt (ref. 60) | 75.4 | 165 | 5.33 |
| expt (ref. 61) | 75.6 | 156 | 5.81 |
| expt (ref. 62) | 75.7 | 163.4 | 5.38 |
| LAPW-GGA (ref. 24) | 69.0 | 292 | 4.4 |
| PAW-GGA (ref. 59) | 69.2 | 293 |  |
| LAPW-LDA (ref. 24) | 64.7 | 344 | 4.4 |
| afmI, LAPW-GGA (ref. 24) | 70.5 | 210 | 5.5 |
| afmII, LAPW-GGA (ref. 24) | 71.2 | 209 | 5.2 |